\documentclass[twocolumn,prd,showpacs,10pt]{revtex4}
\usepackage{graphicx}
\usepackage{dcolumn}
\usepackage{bm}

\newcommand{\square}{\kern1pt\vbox{\hrule height 1.2pt\hbox{\vrule
width 1.2pt\hskip 3pt
\vbox{\vskip 6pt}\hskip 3pt\vrule width 0.6pt}\hrule
height 0.6pt}\kern1pt}

\newcommand{\be}{\begin{equation}}
\newcommand{\ee}{\end{equation}}

\begin{document}

\title{Dynamics of a scalar field in Robertson-Walker spacetimes}

\author{Edmund J. Copeland\footnote{ed.copeland@nottingham.ac.uk} $^{1}$, 
Shuntaro Mizuno\footnote{shuntaro.mizuno@nottingham.ac.uk} $^{1,2}$, and 
Maryam Shaeri\footnote{ppxms1@nottingham.ac.uk } $^{1}$}
\affiliation{\vspace{.5cm}\\ $^1$ School of Physics and Astronomy, University of Nottingham, University Park, Nottingham NG7 2RD, UK \\  
$^2$ Research Center for the Early Universe (RESCEU), School of Science, University of Tokyo, 7-3-1 Hongo, Bunkyo, Tokyo~113-0033, Japan}

\begin{abstract}
We analyze the dynamics of a single scalar field in Friedmann-Robertson-Walker
universes with spatial curvature. We obtain the fixed point solutions
which are shown to be late time attractors. In particular, we determine 
the corresponding scalar field potentials which correspond to these 
stable solutions. The analysis is quite general and incorporates expanding 
and contracting universes with both positive and negative scalar potentials. 
We demonstrate that the known power law, exponential, 
and de-Sitter solutions are certain limits of our general set of solutions.
\end{abstract}

\vskip 1pc 
\maketitle

\section{Introduction}

Scalar fields have played a very important role
in modern cosmology. Today's observed acceleration of the universe,
for example, may be explained by 
the dynamics of a scalar field (for a review, see \cite{Copeland:2006wr}).
The scenarios proposed to solve
the initial conditions of the standard Big Bang theory
such as inflation \cite{Linde,Lyth:1998xn}, 
pre-big bang \cite{Gasperini:1992em,Lidsey:1999mc,Gasperini:2002bn}, and the 
ekpyrotic/cyclic \cite{Khoury:2001wf,Kallosh:2001ai,Steinhardt:2001st} scenarios
also usually require a scalar field.
What makes these scenarios interesting is 
the existence of attractor solutions which implies that
the dynamical system becomes insensitive to the initial conditions. 
A class of such attractor solutions is referred to as 
scaling solutions, where the energy density can be divided into 
contributions which scale with one another throughout the evolution of the universe. 
Investigating the nature of scaling solutions allows one to understand the asymptotic behaviour of
a particular cosmology and helps determine whether such behaviour
is stable or just a transient feature.

In a spatially flat Friedmann-Robertson-Walker (FRW) 
universe filled with a perfect fluid 
and a canonical scalar field,
such scaling solutions are obtained through a simple exponential potential \cite{Copeland:1997et}.
The attractor behaviour in this system has been analyzed extensively
\cite{Lucchin:1984yf,Ratra_Peebles,trac,Ferreira:1997hj,Heard:2002dr,Padmanabhan:2002cp}.
More recently they have been obtained for a wide class of 
modified cosmologies proposed by fundamental theories. 
The cosmologies inlcude those with a non-minimally coupled scalar field 
\cite{Uzan:1999ch,Amendola:1999qq,Holden:1999hm,Gumjudpai:2005ry,Pettorino:2005pv,Amendola:2006qi},
braneworld cosmologies
\cite{Huey:2001ae,Huey:2001ah,Mizuno:2002wa,Savchenko:2002mi,Mizuno:2004xj,Sami:2004rk},
tachyon cosmologies
\cite{Padmanabhan:2002sh,Feinstein:2002aj,Aguirregabiria:2004xd,Piazza:2004df,Copeland:2004hq,
Gumjudpai:2006hg,Tsujikawa:2006mw,Gong:2006sp,Martin:2008xw,Guo:2008sz,Sen:2008yt,Gumjudpai:2009uy},
loop quantum cosmology,
\cite{Lidsey:2004uz,Copeland:2005xs,Copeland:2007qt}, 
phantom cosmology
\cite{UrenaLopez:2005zd} and 
Gauss-Bonnet cosmology
\cite{Tsujikawa:2006ph, Uddin:2009wp} 
(For examples of similar analysis in the context of modified gravity, 
see \cite{Amendola:2006we,Li:2007xn,Fay:2007gg,Amendola:2007nt,Carloni:2007br,Tsujikawa:2007xu,Tsujikawa:2008uc,DeFelice:2008wz,Zhou:2009cy}).  

Fundamental theories may introduce modifications to the 
Friedmann equation in various models that may  or may not be of a similar form. 
Since the stability of the background cosmology is understood to be 
a desirable property, the pragmatic approach is to accommodate 
such modifications in the phase space analysis of the standard 
dynamical system. For the case of a flat FRW universe, 
the analytical method of obtaining the scaling solutions
has been extended to a general set-up, where
the Hubble parameter is given either by an arbitrary power 
\cite{Tsujikawa:2004dp,Calcagni:2004wu}
or an arbitrary function  \cite{Copeland:2004qe}
of the total energy density. 
 
According to the latest results of WMAP5
\cite{Komatsu:2008hk}, although limits being placed
on the flatness of our observable universe are sqeezing the possible outcomes, 
$-0.063 < \Omega_k < 0.017$, there is still the possibility that
we live in a spatially curved FRW universe. It is appropriate, 
therefore, to 
investigate the consequences of curvature on the 
stability of the background cosmology.

In this paper, using the method developed in
\cite{Copeland:2004qe} in the context of braneworld 
corrections made to the standard cosmology,
we will obtain the scaling solutions in spatially 
curved FRW universes and classify the asymptotic
behaviour of the systems.
We aim to present a general approach to analysing 
such systems by deriving the form of the scalar potential 
leading to late time attractors. In doing so, we will demonstrate that
our results reduce to the known results in the literature once
applied to the corresponding models \cite{Halliwell:1986ja,vdHCW}-
(considerable attention has also been given to scaling solutions in Bianchi spacetimes
\cite{Burd:1988ss,Kitada:1992uh,Coley:1997nk,Billyard:1998hv,Billyard:1999ia,Coley:2000yc}).
As was done in \cite{Heard:2002dr} for the case of a flat FRW universe,
we consider positive and negative potentials of the scalar field,
since negative potentials can provide interesting
cosmological scenarios both in an expanding 
\cite{Linde:2001ae,Felder:2002jk,Copeland:2007qt} and 
collapsing universe 
\cite{Khoury:2001wf,Kallosh:2001ai,Steinhardt:2001st}.

The rest of the paper is arranged as follows. 
In Sec.~II we present the equations of motion
and introduce the variables which allow the scaling
solutions to be determined. After analyzing the
stability of these solutions in Sec.~III, we obtain
the general relations which hold when the scaling
solutions are obtained in Sec.~IV. Then, we apply
these results to the open and the closed FRW universe
in Sec.~V and Sec.VI, respectively. Finally, we summarise in Sec.~VII.

\section{Equations of motion}

We consider Friedmann-Robertson-Walker (FRW) cosmologies
such that the dynamics is determined by an effective 
Friedmann equation of the form

\be
\label{b_eq}
H^2 = \frac{8 \pi}{3 m_4^2} \rho L^2(\rho(a))\,,
\ee
where $H \equiv \dot{a}/a$ is the Hubble parameter, $a$
is the scale factor, $\rho(a)$ is the total energy density of the universe, 
a dot denotes differentiation with respect to cosmic 
time and $m_4$ is the four--dimensional Planck mass. 
Throughout this paper, we work in units where $8 \pi/m_4^2=1$,
and we introduce the notation $H = \epsilon \sqrt{H^2}$, where
$\epsilon = \pm 1$ corresponds to an expanding or a contracting 
universe, respectively. Modifications to standard relativistic cosmology are 
parameterized by the function $L(\rho(a))$ and this is assumed to be positive--definite 
without loss of generality. 

We will investigate models where the universe is   
sourced by a self--interacting scalar field $\phi$ with 
potential $V(\phi )$ together with a barotropic fluid 
with equation of state $p_{\gamma} = (\gamma - 1) \rho_{\gamma}$,
where $\gamma$ is the adiabatic index. The energy density and pressure of 
the
scalar field are given by $\rho_\phi = \dot{\phi}^2/2+V$ and 
$p_{\phi} = \dot{\phi}^2/2 -V$, respectively. 
We note here that the effective adiabatic index
of the scalar field is given by 
$\gamma_\phi = (\rho_\phi + p_\phi)/\rho_\phi$.
As in conventional 
cosmologies, we assume that the energy--momenta of these matter 
fields are covariantly conserved and this implies that 

\begin{eqnarray}
\label{fluidconserve}
\dot{\rho}_{\gamma} &=& - 3\gamma H \rho_{\gamma},\\
\label{scalareom}
\ddot{\phi} &=& -3H\dot{\phi} - V_{,\phi}\,,
\end{eqnarray}
where the subscript $\phi$ means differentiation with respect to the field.
Eqs. (\ref{b_eq})--(\ref{scalareom}) close the system that determines 
the cosmic dynamics. Introducing the variables \cite{Huey:2001ae,Huey:2001ah}

\be
\label{defX&Y}
X \equiv \frac{\dot{\phi}}{\sqrt{2\rho}}\,, \hspace{1cm} Y \equiv \frac{\sqrt{|V|}}{\sqrt{\rho}}\,,
\ee
where $\rho$ is the total energy density of the universe, and we adopt the notation
$V = \alpha |V|$ for $\alpha= \pm1$ for positive and negative potentials, respectively. 
Eqs. (\ref{b_eq})-(\ref{scalareom}) can be rewritten in the form

\begin{eqnarray}
\label{b_eq_x}
X_{,N} &=& -3X + \epsilon \alpha \lambda \sqrt{\frac{3}{2}} Y^2 
+ \frac{3}{2} X [2 X^2 + \gamma (1- X^2 - \alpha Y^2)]\,,\nonumber\\
\\
\label{b_eq_y}
Y_{,N} &=& -\epsilon \lambda \sqrt{\frac{3}{2}} XY 
+ \frac{3}{2}Y[2X^2 + \gamma (1- X^2 - \alpha Y^2)]\,,\nonumber\\
\\
\label{b_eq_l}
\lambda_{,N} &=& - \epsilon \sqrt{6} \lambda^2 (\Gamma -1) X
+3 \lambda [2X^2 + \gamma (1-X^2- \alpha Y^2)]  \nonumber\\
&& \rho \frac{\partial \ln L}{\partial \rho} \,,\nonumber\\
\end{eqnarray}
where $N \equiv \ln a$, the subscript $N$ denotes differentiation with respect to
this parameter, and

\begin{eqnarray}
\label{def_lambda&Gamma}
\lambda \equiv - \frac{1}{L}\frac{V_{,\phi}}{V} \hspace{1cm} \Gamma
\equiv V \frac{V_{,\phi \phi}}{(V_{,\phi})^2}\,.
\end{eqnarray} 

For this new set of variables, the definition of the total energy density implies the constraint
equation

\be
\label{rho_total}
X^2 + \alpha Y^2 +\frac{\rho_{\gamma}}{\rho} = 1\,.
\ee

From Eq. (\ref{fluidconserve}) the scale factor, $a$, is expressed
in terms of the fluid energy density, $\rho_\gamma$, as

\be
\label{rel_a_rho_g}
a=a_{(i)} \rho_{\gamma\;(i)}^{\frac{1}{3 \gamma}} 
\rho_{\gamma}^{-\frac{1}{3 \gamma}}\,,
\ee
where the quantities with subscript $(i)$ are evaluated 
at some initial time.

Differentiating $\lambda$, given by Eq. (\ref{def_lambda&Gamma}), with respect to the scalar field, $\phi$,
for the special case of $\lambda$ being a constant, one can show that

\be
\label{GammaConstraint}
\Gamma = 1+ \frac{d\ln L}{d \ln |V|}\,.
\ee

This is the case we consider for the rest of the paper.

\section{Stability}

In order to carry out the stability analysis for this case, it is sufficient to solve Eqs. (\ref{b_eq_x})-(\ref{b_eq_y})
for variables $X$ and $Y$, since Eq. (\ref{b_eq_l}) reduces to the constraint equation (\ref{GammaConstraint}).
For this system of equations we find the following set of physical fixed points

\begin{eqnarray}
\label{kinetic}
X_c &=& \pm 1 \hspace{1cm} Y_c = 0\,, \\ 
\label{fluid}
X_c &=& 0 \hspace{1cm} Y_c = 0\,, \\ 
\label{fluid-scalar-scaling}
X_c &=& \sqrt{\frac{3}{2}}\frac{\gamma}{\lambda } \, \epsilon \hspace{1cm} Y_c =  \sqrt{\frac{3(2-\gamma) \gamma}{2\lambda^2 }\,\alpha }\,,\\ 
\label{scalar-scaling}
X_c &=& \frac{\lambda \epsilon}{\sqrt{6}} \hspace{1cm} Y_c = \sqrt{ \left( 1-\frac{\lambda^2}{6} \right) \alpha }\,.
\end{eqnarray}

In what follows we describe the classification of these solutions in terms of 
their region of existence and stability for an expanding universe containing 
a scalar field with a positive potential (i.e. $\epsilon =1$, and $\alpha =1$).
Later, we will generalise our analysis for an expanding or a contracting universe
in which the scalar field could have either a positive or a negative potential.
These are summarised in Table \ref{StabilitySummary}.

The first two points in (\ref{kinetic}) correspond to the scalar field kinetic energy
dominated solutions. 
The third point (\ref{fluid}) corresponds to the fluid dominated solution.
Upon imposing the constraint Eq. (\ref{rho_total}), 
the fourth point (\ref{fluid-scalar-scaling}) is a solution that exists
for $\lambda \neq 0$, $0 < \gamma < 2$, and $\lambda^2 > 3\gamma$; and it describes a 
scenario where, for a given fluid, the contribution of the scalar field density to the total energy density
scales with that of the fluid to the total density. i.e. $X_c^2 + \alpha
Y_c^2 = \frac{3\gamma}{\lambda^2}$. 
For convenience, we refer to this solution as the fluid-scalar
field scaling solution throughout this paper. The final point (\ref{scalar-scaling}) arises if 
$\lambda^2 < 6$, and the constraint Eq. (\ref{rho_total}) implies that, in this case, the energy density of the universe 
is dominated by the scalar field, having an effective adiabatic index $\gamma_\phi = \frac{\lambda^2}{3}$.
This describes a scaling solution, where as the universe evolves, the kinetic energy and the potential
energy of the scalar field scale together. 
We refer to this solution as the scalar field dominated scaling solution
throughout this paper.

Having obtained the scaling solutions, we need to 
investigate their stability to small fluctuations. 
Considering perturbations of the form
\be
X = X_c + \delta X\,, \hspace{1cm} Y = Y_c + \delta Y\,,
\ee
where $\delta X \propto e^{w N}$, and $\delta Y \propto e^{w N}$,
and expanding Eqs. (\ref{b_eq_x})-(\ref{b_eq_y}), for the kinetic energy dominated
solutions, 
\be
\label{}
w_+ = 3(2-\gamma)  \hspace{1cm } w_- = 3 \mp \sqrt{\frac{3}{2}} \lambda \epsilon \,, 
\ee
respectively. In this case, stability is only achieved for fluids with adiabatic index of
$\gamma>2$, which is not satisfied for known realistic fluids. These points,
are therefore considered to be unstable. For the fluid dominated scenario, 
the eigenvalues are given by 

\be
\label{}
w_+ = \frac{3\gamma} {2}  \hspace{1cm } w_- = -3+\frac{3\gamma}{2} \,,
\ee
which is clearly only stable for negative values of the adiabatic index, $\gamma$.
We will therefore consider this solution to be unstable for all realistic types of
fluid. The fluid-scalar field scaling solution
yields the eigenvalues (recall $0<\gamma<2$, and $\lambda^2 > 3\gamma$)

\be
\label{}
w_\pm = \frac{3}{4} (\gamma-2) \left( 1 \pm \sqrt{ 1 - \frac{8 \gamma (3\gamma - \lambda^2)}{\lambda^2 (\gamma-2)}   }  \right)\,,
\ee
which implies unconditional stability when these solutions exist.
The same analysis yields the following eigenvalues for 
the scalar field dominated scaling solutions

\be
\label{}
w_+ =\frac{1}{2} (\lambda^2-6)  \hspace{1cm } w_- = \lambda^2 - 3\gamma \,, 
\ee
which indicates that when these solutions exist, as long as $\lambda^2 < 3\gamma$, stability is guaranteed.
We also note that in a contracting universe ($\epsilon=-1$), an attractor solution correpsonds to one with {\em positive}
real eigenvalues. This is because the parameter with respect to which the dynamical 
system of (\ref{b_eq_x})-(\ref{b_eq_y}) is described, $N$, becomes a decreasing function of time. 
Keeping this in mind, the general consideration of various combinations of an expanding/contracting universe containing
a scalar field with a positive/negative potential are captured in Table {\ref{StabilitySummary}} below. 
For the case of $L = 1$ (i.e. a flat FRW universe), a similar classification was done for simple 
exponential potentials in \cite{Heard:2002dr}.

\begin{widetext}
\begin{center}
\begin{table} [h!]
\begin{tabular} { |c|c|c|c|c|c|c|c|c|c|c| }
\hline
\raisebox{1.5ex} { } & \multicolumn{2}{|c|}{$X=1$} & \multicolumn{2}{|c|}{$X=-1$} & \multicolumn{2}{|c|}{$X=0$} & \multicolumn{2}{|c|}{$X=\sqrt{\frac{3}{2}} \frac{\gamma}{\lambda}\epsilon$} & \multicolumn{2}{|c|}{$X=\frac{\lambda \epsilon}{\sqrt{6}}$}  \\ 
 & \multicolumn{2}{|c|}{$Y=0$} & \multicolumn{2}{|c|}{$Y=0$} & \multicolumn{2}{|c|}{$Y=0$} & \multicolumn{2}{|c|}{$Y=\sqrt{\frac{3(2-\gamma)\gamma}{2\lambda^2} \, \alpha}$} & \multicolumn{2}{|c|}{$Y=\sqrt{\left( 1-\frac{\lambda^2}{6} \right) \, \alpha}$}  \\
\hline 
 & exists & stable & exists & stable & exists & stable & exists & stable & exists & stable \\
\hline
$\epsilon=+1$ \,, $\alpha=+1$ & $\forall \lambda, \, \forall \gamma$ & No & $\forall \lambda, \, \forall \gamma$ & No & $\forall \lambda, \, \forall \gamma$ & No & $\lambda \neq 0, \,\, \lambda^2>3\gamma$ & when exists & $\lambda^2<6$ & $\lambda^2<3\gamma$\\      
\hline
$\epsilon=+1$ \,, $\alpha=-1$ & $\forall \lambda, \, \forall \gamma$ & No & $\forall \lambda, \, \forall \gamma$ & No & $\forall \lambda, \, \forall \gamma$ & No & No & - & $\lambda^2>6$ & No\\      
\hline
$\epsilon=-1$ \,, $\alpha=+1$ & $\forall \lambda, \, \forall \gamma$ & $0>\lambda>-\sqrt{6}$ & $\forall \lambda, \, \forall \gamma$ & $0<\lambda<\sqrt{6}$ & $\forall \lambda, \, \forall \gamma$ & No & $\lambda \neq 0, \,\, \lambda^2>3\gamma$ & No & $\lambda^2<6$ & No\\      
\hline
$\epsilon=-1$ \,, $\alpha=-1$ & $\forall \lambda, \, \forall \gamma$ & $0>\lambda>-\sqrt{6}$ & $\forall \lambda, \, \forall \gamma$ & $0<\lambda<\sqrt{6}$ & $\forall \lambda, \, \forall \gamma$ & No & No& - & $\lambda^2>6$ & when exists \\      
\hline
\end{tabular}
\caption{This table summarises the existence and stability conditions for an expanding ($\epsilon=1$) or a contracting ($\epsilon=-1$)
universe containing a fluid with the adiabatic index $\gamma$ and a scalar field with either a positive ($\alpha=1$) or a negative ($\alpha=-1$)
potential. }
\label{StabilitySummary}
\end{table}
\end{center}
\end{widetext}

\section{General relations for Scaling Solutions}

In the remainder of this paper we concentrate on 
the solutions
given by 
Eqs. (\ref{fluid-scalar-scaling})-(\ref{scalar-scaling}).
We aim to derive the scalar field potentials which correspond to these late time attractors for particular forms of $L(\rho)$
based on spatially curved cosmologies. Working at the fixed points,
from $Y_c = \sqrt{\frac{|V|}{\rho}}$, and the fact that $\rho_\phi/\rho = X_c^2 + \alpha Y_c^2 =$ constant for these solutions, $L(\rho)$
can be described as $L(V)$. Moreover, from the definition of $\lambda$, one can show that

\be
\label{V_L_phi}
\int{\frac{dV}{V L}} = - \lambda \phi\,.
\ee

Given a specific form of $L$, therefore, it is possible to derive the scalar potential 
resulting in the scaling solution by integrating Eq. (\ref{V_L_phi}).  We note that, as 
demonstrated in \cite{Copeland:2004qe}, Eq. (\ref{V_L_phi}) is equivalent to imposing 
the constraint Eq.~(\ref{GammaConstraint}) throughout the evolution of the field.

Furthermore, since at the fixed points, $X=X_c$, $\phi$ is a monotonically varying function of 
cosmic time, $t$, and can be considered as a suitable dynamical variable for the system.
 We note here that this assumption is only valid when $X_c \neq 0$.
In general, the scalar field Eq. (\ref{scalareom}) can be expressed in the form

\be
\dot{\rho}_\phi = -3H \dot{\phi}^2\,,
\ee
and for the scaling solutions, this equation can be expressed as

\be
\label{rel_rho_ph}
\dot{\phi} = - \alpha \epsilon \frac{1}{Y_c \sqrt{3}} \left(\frac{\rho_\phi}{\rho}\right) \frac{1}{L(V)} \frac{V_{,\phi}}{\sqrt{|V|}}\,,
\ee
but since $V$ and $\phi$ are related through Eq. (\ref{V_L_phi}), this
can be integrated to find $t$ as a function of the scalar field

\be
\label{rel_t_ph_gen}
t = - \alpha \epsilon Y_c\sqrt{3} \left(\frac{\rho}{\rho_\phi} \right) \int^\phi d \phi L(V) \frac{\sqrt{|V|}}{V_{,\phi}} \,.
\ee

\section{Open FRW universe}

Here, we consider what form of the scalar field potential provides
the fixed point solution characterised by $(X,Y) = (X_c, Y_c)$ in 
the open FRW universe. In this case, $L(\rho)$ is given by

\be
\label{L_open_FRW}
L(\rho) = \sqrt{1 + \frac{3}{\rho a^2}}\,.
\ee

In our analysis, we assume that $\lambda \neq 0$,
but we return to this point and consider the special case of $\lambda = 0$ at
the end of this section. We note at this point that an expanding (contracting)
open universe obeying the Friedmann equation (\ref{b_eq}) could stop its expansion
(contraction) process and begin to contract (expand) if the scalar field has a
negative (positive) potential. It is, therefore, interesting to study the physics 
obtained from different combinations of $\epsilon$ and $\alpha$ in an open universe.
We begin our discussion by concentrating first on an expanding ($\epsilon=1$)
universe sourced by a positive potential ($\alpha=1$) scalar field.

\subsection{Case A: Fluid-scalar field scaling solutions} \label{Case A}

When the universe is expanding ($\epsilon=1$), the solution given by (\ref{fluid-scalar-scaling})
exists and is stable for $\lambda^2 > 3 \gamma$, 
and the scale factor $a$ can be expressed in this case as 

\be
\label{a_as_rhog}
a = A \rho_\gamma^{-\frac{1}{3\gamma}}\,,\;\; {\rm with}\;\;
A = a_{(i)} \rho_{\gamma(i)}^{\frac{1}{3\gamma}}
\left( \frac{\lambda^2}{\lambda^2 - 3 \gamma}
\right)^{\frac{1}{3\gamma}}\,.
\ee
The correction function $L$ given by Eq. (\ref{L_open_FRW}) can then be rewritten as

\be
\label{L_open_FRW_scaling}
L(\rho) = \sqrt{1 + \frac{3}{A^2} \rho^{\frac{2-3\gamma}{3\gamma}}}.
\ee

By considering the fixed point, substituting this into Eq. (\ref{V_L_phi})
and integrating, yields the scaling solution potential 

\be
\label{V_open_FRW_scaling}
|V(\phi)|^{\mu} = Y_c^{2 \mu}
 \left(\frac{A^2}{3}\right) 
{\rm cosech}^2 \left( \frac{\lambda}{2} \mu \phi \right)\,,
\ee
where $\mu = (2-3\gamma)/3\gamma$.
Notice that the valid range of $\mu$, for the region of 
existence of these solutions, is $-2/3 < \mu < 0$
and $0 < \mu < \infty$. The special case of $\mu = 0$, where
Eq.~(\ref{V_open_FRW_scaling}) is no longer valid, is discussed later.

Using the definition of $Y$ at the fixed point, and substituting Eq. (\ref{V_open_FRW_scaling}) 
into Eq. (\ref{L_open_FRW_scaling}), the correction function can be written in terms of 
the scalar field as $L(\phi)$. Upon substituting this form back in Eq. (\ref{rel_t_ph_gen}) the
time dependence of the scalar field can be evaluated as

\be
\label{rel_t_ph_open_sca}
t= \epsilon \frac{ \sqrt{3} \lambda}{3 \gamma} 
\left(\frac{3}{A^2}\right)^{\frac{1}{2\mu}} \int^\phi d \phi \sinh^{1/\mu} \left( \frac{\lambda \mu}{2}\phi\right)\,.
\ee

Classifying the behaviour of the evolution equations above in terms of the sign 
of the parameter $\mu$, and recognising the sign of the argument inside the brackets
remains invariant under the transformation $\lambda \to -\lambda$ and $\phi \to -\phi$, 
we choose to work in the first quadrant without loss of generality. 

For $\mu>0$, we find that at early times, as $\phi \to 0$,  
the asymptotic form of the potential relating to the 
curvature dominated universe is a power law function 
$V \propto \phi^{-\frac{2}{\mu}}$. However,
at late times, as $\phi \to \infty$, the potential of a fluid-scalar 
field dominated universe is of an exponential form 
$V \sim \exp[ - \lambda \phi]$.

On the other hand, for $\mu<0$, at early times, as $\phi \to \infty$,
where the curvature is negligible, the asymptotic form of the potential
is obtained to be an exponential one $V \sim \exp[ - \lambda \phi]$.
Once the universe becomes dominated by the curvature at late times,
as $\phi \to 0$, the potential can be approximated by a power law function
$V \propto \phi^{-\frac{2}{\mu}}$, recovering the standard result $a \sim t$.

For $\mu=0$, since the correction function 
$L$ is a constant and can be thought of as modification to Newton's 
gravitational constant in Eq.~(\ref{b_eq}), the expansion law becomes that of the flat
Friedmann cosmology, and 
the potential yielding this late time attractor will then be 
the exponential potential as found in \cite{Copeland:1997et,Padmanabhan:2002cp}. We further note that this 
scenario refers to the special case where the contribution of the scalar field, 
the fluid, and the curvature to the Friedmann equation scale together.

We note, according to the results illustrated in Table \ref{StabilitySummary},
that the fixed points (\ref{fluid-scalar-scaling}) do not exist for negative potentials. This is 
true for an expanding or a contracting universe. Although in an expanding (contracting ) universe, 
a negative (positive) scalar potential could slow down the growth of the scale factor and cause 
the universe to collapse (expand), we do not expect the late time attractors to be given by 
these scaling solutions. We can also see from Table \ref{StabilitySummary} that this 
set of solutions does not exist in a contracting universe sourced by a negative 
scalar field potential.

\subsection{Case B: Scalar field dominated scaling solution}

When the universe is expanding ($\epsilon=1$)
the solution given by (\ref{scalar-scaling}),
is an attractor for $\lambda^2<3\gamma$,
and the scalar field dominates over the fluid, resulting in the effective adiabatic index 
$\gamma_{\phi} = \frac{\lambda^2}{3}$. The scale factor is then given by
\be
\label{a_as_rhophi}
a = B \rho^{-\frac{1}{\lambda^2}}\,,\;\;\; {\rm with}
\;\;
B=a_{(i)} \rho_{(i)} ^{\frac{1}{\lambda^2}}\,.
\ee

The form of the correction function $L$ given by
Eq. (\ref{L_open_FRW}) can then be rewritten as

\be
\label{L_open_FRW_p_l}
L(\rho) = \sqrt{1 + \frac{3}{B^2} 
\rho^{\frac{2-\lambda^2}{\lambda^2}}}.
\ee

By considering the attractor solution, substituting this form of the correction function into Eq. (\ref{V_L_phi}),
and integrating, yeilds the scaling solution potential 

\be
\label{V_open_FRW_p_l}
|V(\phi)|^{\nu} = Y_c^{2 \nu} \left(\frac{B^2}{3}\right) {\rm cosech}^2 \, \left( \frac{\lambda}{2}\nu \phi \right)\,,
\ee
where $\nu = (2-\lambda^2)/\lambda^2$, and the form of the potential is
valid for $\lambda^2 \neq 2$ .

By using the definition of $Y$ at the fixed point, 
and substituting
Eq.~(\ref{V_open_FRW_p_l}) into Eq.~(\ref{L_open_FRW_p_l})
the correction function can be written in terms of the scalar field as $L(\phi)$. From this form
the time dependence of the scalar field can be evaluated using Eq.(\ref{rel_t_ph_gen}) to be

\be
\label{rel_t_ph_open_pow}
t= \epsilon \frac{\sqrt{3}}{\lambda} 
\left(\frac{3}{B^2}\right)^{\frac{1}{2\nu}} \int^\phi d \phi \sinh^{1/\nu} \left(\frac{\lambda \nu}{2} \phi\right)\,,
\ee

By analogy, the asymptotic behaviour of the universe in this scenario can be obtained 
using the same method as employed in {\em Case A}. These results are summarised in Table \ref{Open_FRW}.

\begin{widetext}
\begin{center}
\begin{table} [h!]
\begin{tabular} { |c|c|c|c|c|c|c| }
\hline
 & \multicolumn{3}{|c|}{\em Case A} & \multicolumn{3}{|c|}{\em Case B} \\
\hline
 & $\mu<0$ & $\mu=0$ & $\mu>0$ & $\nu<0$ & $\nu=0$ & $\nu>0$ \\
\hline \hline
 & $V \sim \exp[ - \lambda \phi]$ & $V \sim \exp[ - 2 \phi]$ & $V \sim \phi^{-\frac{2}{\mu}}$ & $V \sim \exp[ - \lambda \phi]$ & $V \sim \exp[ - 2 \phi]$ & $V \sim \phi^{-\frac{2}{\nu}}$   \\
\raisebox{1.5ex} {Early times} & negligible & scaling & curvature & negligible & scaling & curvature \\ 
 & curvature & curvature & dominated & curvature & curvature & dominated \\
\hline
 & $V \sim \phi^{-\frac{2}{\mu}}$ & $V \sim \exp[ - 2 \phi]$ & $V \sim \exp[ - \lambda \phi]$ & $V \sim \phi^{-\frac{2}{\nu}}$ & $V \sim \exp[ - 2 \phi]$ & $V \sim \exp[ - \lambda \phi]$  \\
\raisebox{1.5ex} {Late times} & curvature & scaling & negligible & curvature & scaling & negligible \\ 
 & dominated & curvature & curvature & dominated & curvature & curvature \\
\hline
\end{tabular}
\caption{This table summarises the asymptotic behaviour of an expanding ($\epsilon = 1$) open FRW universe described by the 
scaling solutions, when the scalar potential is positive. 
{\em Case A} refers to the fluid-scalar field scaling solution, and {\em Case B} 
corresponds to the scalar field dominated 
scaling solution.}
\label{Open_FRW}
\end{table}
\end{center}
\end{widetext}

We now turn to our general set of results in Table \ref{StabilitySummary} and note that 
in a contracting universe ($\epsilon=-1$), 
if the scalar potential is negative ($\alpha=-1$)
and steep enough ($\lambda^2 > 6$), 
the scalar field dominated scaling solution
can also result in stable attractor behaviour. 
Such solutions will be unconditionally stable if they exist, 
and we find
the form of the potential together with 
the time evolution of the scalar field are still given by Eqs.~(\ref{V_open_FRW_p_l})
 and (\ref{rel_t_ph_open_pow}).
 In this scenario, $\nu$ is clearly negative, and due to the time reversal of
Eq.~(\ref{rel_t_ph_open_pow}) for $\epsilon =-1$, one expects from Table \ref{Open_FRW},
an early time power law behaviour of the potential to be followed by an exponential form
at late times. This is the generalisation of \cite{Heard:2002dr} to general curved space scenarios.

\subsection{Case C: $\lambda \approx 0$} 

We now return to the special case of $\lambda = 0$, and notice that this
corresponds to a constant potential (i.e. de-Sitter space). Solutions
for this type of universe are known \cite{deSitter}, 
and it is therefore interesting to 
see how they would fit into the larger class of solutions we are presenting here.
Considering the scaling solutions, we notice that the fixed points described by
(\ref{fluid-scalar-scaling}) do not exist for the class of constant potentials. However,
the attractor solutions (\ref{scalar-scaling}) reduce to $X_c =0$, and $Y_c =1$, which describe 
an exact de sitter solution. As mentioned previously, the argument of using $\phi$
as a monotonically varying function of time breaks down when $X_c \propto \dot{\phi} = 0$.
For this simplified case, one can clearly solve Eqs. (\ref{b_eq})-(\ref{scalareom}) directly to find the 
exact solution. We notice, however, that we should also be able to recover this solution 
by investigating the behaviour of the scalar potential and the scale factor 
very close to the fixed point. We do this by considering small values of $\lambda$.
From Eq. (\ref{rel_t_ph_open_pow}), we find the asymptotic dependence of the scalar field 
to be given by $t \propto \phi$. Substituting this into Eq. (\ref{V_open_FRW_p_l}), and using 
Eq. (\ref{rel_a_rho_g}) near the fixed point, we find that the scale factor evolves as
$a \propto \sinh \left( \frac{\lambda}{2} \nu t \right)$. Thus, recovering the de-sitter solution as a
special case of our larger set of solutions derived here.

We further note from Table \ref{StabilitySummary} that when $\lambda \approx 0$, a negative potential scaling solution does not exist 
in an expanding or a contracting universe; and the positive potential solution, which exists in a contracting universe, 
is unstable.

\section{Closed FRW universe}

Here, we aim to derive the forms of the scalar potentials which would result
in stable fixed points (\ref{fluid-scalar-scaling})-(\ref{scalar-scaling}), corresponding to the scaling
solutions in a closed FRW universe. In this case, 
$L(\rho(a))$ is given by

\be
\label{L_closed_FRW}
L(\rho) = \sqrt{1 - \frac{3}{\rho a^2}}\,.
\ee

In this scenario, the curvature can not dominate the 
contributions to the total energy density of the universe, otherwise the correction function 
(\ref{L_closed_FRW}) would be imaginary and the right hand side of the Friedmann equation
would become negative. The valid range of the correction function is thus given by $0 \le L \le 1$.
The case of $L=0$ corresponds to a universe with a constant scale factor, and describes the case
where there is no expansion or contraction taking place. The universe may undergo a bounce at 
this point and change from an expanding (contracting) behaviour to a contracting (expanding) one.
We bear this in mind in our following analysis. 
As in the open case, we first concentrate on
the expanding universe ($\epsilon = 1$) with positive
potential ($\alpha =1$) scalar field, and
we will comment on the negative potentials and 
the difference between an expanding
and a contracting universe for each set of scaling solutions.

\subsection{Case A: Fluid-scalar field scaling solution} \label{Case A_closed}

In an expanding universe ($\epsilon=1$), the solution given by (\ref{fluid-scalar-scaling}) 
is an attractor for $\lambda^2 > 3 \gamma$,
and describes a scenario where the contribution of the scalar field scales with that of the fluid. 
The scale factor in this case is also given by Eq.~(\ref{a_as_rhog}), and 
after substituting this into Eq.~(\ref{L_closed_FRW}), 
the form of the correction function, $L$, is found to be 

\be
\label{L_closed_FRW_scaling}
L(\rho) = \sqrt{1 - \frac{3}{A^2} \rho^{\frac{2-3\gamma}{3\gamma}}}.
\ee

Notice there is a maximum level of density, that depends on $A$, beyond which $L$
lies outside its valid range and the situation is unphysical. Following our previous analysis, 
except for the case $\mu=0$, where $\mu = (2-3\gamma)/3\gamma$, the potential yielding the fixed point solution
is given by

\be
\label{V_closed_FRW_scaling}
|V(\phi)|^\mu = Y_c^{2 \mu} \left(\frac{A^2}{3}\right) {\rm sech}^2 \left( \frac{\lambda}{2} \mu \phi \right)\,.
\ee

Using $Y_c = \sqrt{|V|/\rho}$, and substituting Eq.~(\ref{V_closed_FRW_scaling}) into Eq.~(\ref{L_closed_FRW_scaling}),
the correction function can be written as $L(\phi) = \tanh \left( \frac{\lambda}{2} \mu \phi \right)$. 
And substituting this form of $L$ back into Eq.~(\ref{rel_t_ph_gen}),
yields the time dependence of the scalar field as

\be
\label{rel_t_ph_closed_sca}
t= \epsilon \frac{ \sqrt{3} \lambda}{3 \gamma} 
\left(\frac{3}{A^2}\right)^{\frac{1}{2\mu}} \int^\phi d \phi \cosh^{1/\mu} \left(\frac{\lambda \mu}{2}\phi\right)\,.
\ee

In order to examine the asymptotic behaviour of the scalar field potential, we follow our previous
line of argument and divide up the region of validity of $\mu$ 
into its positive and negative values ( i.e. $-2/3 < \mu < 0$, and $0 < \mu < \infty$). 
Once again, making use of the symmetry $\lambda \to -\lambda$ and $\phi \to -\phi$, we
restrict our analysis to the potentials lying in the first quadrant, without loss of generality. 

For $\mu>0$ (i.e. $3\gamma <2$), in an expanding universe 
at early times, since $a^{-2}>a^{-3\gamma}$, one would expect the curvature 
to dominate initially. However, as mentioned above, this can not happen
in a closed universe, otherwise the correction function (\ref{L_closed_FRW_scaling}) 
becomes imaginary and consequently, the right hand side of the Friedmann Eq.~(\ref{b_eq}) 
will be negative.
A pragmatic starting point for the expanding evolution 
is, therefore, where the curvature contribution is only just subdominant to the
scalar field and the fluid densities. We set this point to correspond to $t \approx 0$, since 
going back in time from this point, in the absence of motivations from a fundamental theory,
is not physically meaningful. This is when the total energy density approaches
its maximum value $\rho \to \rho_{max} \equiv (A^2/3)^{1/\mu}$
and so, the correction function $L(\phi)\to 0$. Eq.~(\ref{b_eq}) then suggests that 
the scale factor does not change as $H \to 0$. Once the universe starts expanding,
the contribution of the fluid and the scalar field to the energy density starts to 
dominate that of the curvature. In the limit where the curvature has become negligible and the 
universe asymptotes to a flat FRW spacetime, we recover the exponential limit of the potential 
$V \sim \exp [-\lambda \phi]$.

For $\mu<0$, as the 
universe expands, it starts from an asymptotically flat FRW state dominated 
by the fluid and the scalar field. The potential in this case is that of the exponential 
form $V \sim \exp [-\lambda \phi]$, and the curvature can be neglected. In time, we 
reach the maximum level of energy density, where the contribution of the curvature 
becomes very close to that of the combination of the scalar field and the fluid, and 
the scale factor seizes to grow as $H \to 0$. 
After this turning point, the universe begins to collapse, and
this provides us with a natural motivation to consider a contracting
universe. However, one can see from Table \ref{StabilitySummary},
that in a contracting universe, late time fluid-scalar field scaling attractor solutions either do not
exist at all (for $\alpha =-1$) or are unconditionally unstable when they
do exist (for $\alpha =1$). Once the universe starts to collapse, 
solutions will asymptote towards the kinetic dominated solutions for $\lambda^2<6$. 
On the other hand, for $\lambda^2>6$, solutions will asymptote towards
the fluid dominated scaling solutions if the potential is negative ($\alpha = -1$), 
and will be unstable for positive potentials.

It is worth noting that in the limit of $H \to 0$, as a result of $L \to 0$,
the total energy density is a constant close to its maximum value. 
From Eqs.~(\ref{defX&Y}) and (\ref{V_closed_FRW_scaling}), 
we conlcude that in this limit the potential is 
almost constant $V \approx Y_c^2 \left( \frac{A^2}{3} \right)^{1/\mu}$.
One may then naively expect the universe to expand exponentially as 
it would do in a de-sitter case; however, we note that the correction factor
stops this behaviour by keeping the scale factor constant in
Eq.~(\ref{b_eq}).

The special case of $\mu=0$ corresponds to where the contribution of the curvature,
the fluid, and the scalar field are all scaling at all times during the evolution. We note
that this is only valid if the curvature does not dominate over the combination of the scalar field 
and the fluid, in which case, the correction function of Eq.~(\ref{L_closed_FRW_scaling}) 
is a constant and the potential has the exponential form.

Similarly to the open universe scenario described earlier, 
we can see from Table \ref{StabilitySummary} that
for negative scalar potentials ($\alpha=-1$)
this set of scaling solutions do not exist in either an
expanding universe or a contracting one. 

\subsection{Case B: Scalar field dominated scaling solution} 

In an expanding universe ($\epsilon=1$)
when the potential of the scalar field is positive
($\alpha=1$), the fixed points given by 
(\ref{fluid-scalar-scaling}) are attractor solutions for 
$\lambda^2<3\gamma$, and the scale factor evolves according to Eq. (\ref{a_as_rhophi}). 
Therefore, the form of the correction function
given by Eq. (\ref{L_closed_FRW}) becomes

\be
\label{L_closed_FRW_p_l}
L(\rho) = \sqrt{1 - \frac{3}{B^2} 
\rho^{\frac{2-\lambda^2}{\lambda^2}}}.
\ee

We remember that in this case, the contribution of the fluid is negligible for 
these solutions at all times during the evolution of the universe.

Similarly to our previous analysis, we find the corresponding scalar potential, except for
the case of $\nu=0$, to be given by

\be
\label{V_closed_FRW_p_l}
|V(\phi)|^\nu = Y_c^{2\nu} \left(\frac{B^2}{3}\right) {\rm sech}^2 \left( \frac{\lambda}{2} \nu \phi \right)\,,
\ee
and the time dependence of the scalar field is

\be
\label{rel_t_ph_closed_p_l}
t= \epsilon \, \frac{1}{\sqrt{3} \lambda} 
\left(\frac{3}{B^2}\right)^{\frac{1}{2\nu}} \int^\phi d \phi \cosh^{1/\nu} \left(\frac{\lambda \nu}{2}\phi\right)\,.
\ee

By analogy, the asymptotic behaviour of an expanding universe can be obtained through
a similar discussion as the previous case. These results are summarised in Table \ref{Close_FRW}.

\begin{widetext}
\begin{center}
\begin{table} [h!]
\begin{tabular} { |c|c|c|c|c|c|c| }
\hline
 & \multicolumn{3}{|c|}{\em Case A} & \multicolumn{3}{|c|}{\em Case B} \\
\hline
 & $\mu<0$ & $\mu=0$ & $\mu>0$ & $\nu<0$ & $\nu=0$ & $\nu>0$ \\
\hline \hline
 & $V \sim \exp[ - \lambda \phi]$ & $V \sim \exp[ - 2 \phi]$ & $V \approx Y_c^2 \left( \frac{A^2}{3} \right)^{1/\mu}$  & $V \sim \exp[ - \lambda \phi]$ & $V \sim \exp[ - 2 \phi]$ & $V \approx Y_c^2 \left( \frac{B^2}{3} \right)^{1/\nu}$    \\
\raisebox{0 ex} {Early times} & negligible & scaling & curvature & negligible & scaling & curvature \\ 
 & curvature & curvature & only just & curvature & curvature & only just \\
 & &  & subdominant &  &  & subdominant \\
\hline
 & $V \approx Y_c^2 \left( \frac{A^2}{3} \right)^{1/\mu}$ & $V \sim \exp[ - 2 \phi]$ & $V \sim \exp[ - \lambda \phi]$ & $V \approx Y_c^2 \left( \frac{B^2}{3} \right)^{1/\nu}$ & $V \sim \exp[ - 2 \phi]$ & $V \sim \exp[ - \lambda \phi]$  \\
\raisebox{0 ex} {Late times} & curvature & scaling & negligible & curvature & scaling & negligible \\ 
 & only just & curvature & curvature & only just  & curvature & curvature \\
 & subdominant &  &  &  subdominant & &  \\
\hline
\end{tabular}
\caption{This table summarises the asymptotic behaviour of an expanding ($\epsilon = 1$) closed FRW universe described by the 
scaling solutions, when the scalar potential is positive. {\em Case A}
 refers to the fluid-scalar field scaling solution, 
and {\em Case B} 
corresponds to the scalar field dominated 
scaling solution.}
\label{Close_FRW}
\end{table}
\end{center}
\end{widetext}

It is worth noting at this stage that according to the results demonstrated in Table \ref{StabilitySummary},
an expanding (contracting) universe with a negative (positive) scalar potential does not admit 
attractor solutions. However, for a contracting universe with a negative 
potential, if such solutions exist they will be unconditionally stable, 
and the form of the potential 
together with the time evolution of the scalar field
can be seen to
be given by Eqs.~(\ref{V_closed_FRW_p_l}) and 
(\ref{rel_t_ph_closed_p_l}).

\subsection{Case C: $\lambda \approx 0$} 

As mentioned in the open universe scenario, this case yields an ill-defined solution corresponding to the
fixed points described by (\ref{fluid-scalar-scaling}).
However, $\lambda = 0$ results in a perfectly well-defined scaling solution when substituted in the 
fixed points given by (\ref{scalar-scaling}). The scaling solution then reduces to $X_c = 0$, and $Y_c = 1$, which is clearly a 
de-Sitter solution. Once again, the assumption of $\phi$ being a monotonically varying 
function of time breaks down at $X_c = 0$. Thus,
we take the pragmatic approach of investigating the form of the potential and the behaviour of the scale
factor near the scaling solution by considering small values of $\lambda$. We then find the potential
to be given by $V \propto \cosh^{-2/ \nu} \left( \frac{\lambda}{2} \nu \phi \right)$, and the time dependence
of the scalar field to be $t \propto \phi$, which from Eq. (\ref{a_as_rhophi}), yields the scale factor 
as $a \propto \cosh \left( \frac{\lambda^2}{2} \nu t \right) $. This is the form of the scale factor evolution
in a closed de-sitter universe, and is a special subset of the solutions we have derived here.

In an analogous way to our description for an open scenario, the fixed points (\ref{scalar-scaling}) do 
not exist in an anti-de-Sitter universe; and even though they exist in a contracting universe sourced by a
scalar field with a positive potential, such solutions are unconditionally unstable.

\section{Summary}

In order to understand the asymptotic behaviour of 
a particular cosmology and determine whether such
a background is stable or not, a class of attractor solutions
referred to as the scaling solutions play an important role.
By now, even though many scaling solutions have been
obtained in various cosmologies, they are limited 
to the case of a flat Friedmann-Robertson-Walker
(FRW) universe. In this paper, we have analyzed the dynamics of a 
single scalar field in FRW universes with spatial curvature.
We started by generalising the approach developed in 
\cite{Copeland:2004qe} to incorporate 
expanding and contracting universes filled with a perfect fluid and a 
single scalar field. Due to a growing interest in negative scalar potentials
and the possibility of obtaining them from fundamental theories,
these are also accommodated in our formalism.

We have identified two sets of scaling solutions. One, where the contribution
of the fluid to the total energy density scales with that of the scalar field throughout 
the evolution, and the other, where the scalar field dominates over the fluid, and the kinetic 
energy of the field scales with its potential energy.
By concentrating on both types of scaling solutions, we obtained the generalised 
dynamical  equations. Once a particular form of the modification function is given 
and its dependence on the total energy density is known, these equations can be used to 
derive the form of the scalar potential that leads to the late time attractor solutions.
This analysis is explicitly carried out for the cases of an open and a closed universe. 

After presenting the general form of the potential, we examined the asymptotic 
behaviour of the dynamics. In an open universe sourced by a positive scalar potential, 
we concluded that in regions where the curvature is dominant, the potential can be
approximated by a power law function; and where it is negligible, an exponential form is 
a good approximation. This result is consistent with \cite{vdHCW}, where the exponential 
form of the potential was found to be unstable when the curvature term becomes important.
We have also shown how the well known de-sitter solutions can be found as certain limits 
of our general potential. We then highlighted the case of a collapsing open universe (which
could happen in the presence of a negative potential \cite{Felder:2002jk}) and concluded that, if the potential is 
steep enough, the scalar field dominated scaling solution is a late time attractor. This is in line 
with the known results in the context of an ekpyrotic collapse in the flat FRW universe 
\cite{Khoury:2001wf,Kallosh:2001ai,Steinhardt:2001st}.

The general form of the scalar potential is also presented for a closed universe. Here we 
are facing an interesting scenario and care needs to be taken since the curvature is
forbidden to dominate over the combination of fluid and the scalar field in an expanding universe when 
the potential is positive. In the limit where the curvature is comparable to this combination, 
the potential is a constant and so is the scale factor. When the curvature becomes subdominant, 
the potential asymptotes to an exponential form. In a contracting closed universe we find that
a steep negative potential provides the late time attractor in the form of a scalar field 
dominated solution. Negative exponential potentials have been known to be stable solutions
in ekpyrotic models, and this is a special limit of our derived potential. 

In this work we have concentrated on the scaling solutions, but it is also worth noting that
in a contracting universe, if the potential is flat
enough ($\lambda^2 < 6$), the kinetic dominated solutions
are shown to be late time attractors for both the positive and the 
negative potential scalar fields. This solution
corresponds to the ones obtained in the pre-big bang cosmology 
\cite{Gasperini:1992em,Lidsey:1999mc,Gasperini:2002bn}.

Although the method proposed in \cite{Copeland:2004qe} had been motivated by 
generalising a large class of modifications introduced to the Friedmann equation
in the context of braneworld cosmologies in flat space, we have demonstrated here that
a spatial curvature term can also be written in a similar fashion. We recognise that 
at the background level it is not possible to distinguish between corrections originating from 
fundamental theories that can be encoded within the correction function, $L(\rho)$, and 
information about the spatial curvature which could be represented through a similar 
modification function. In order to differentiate between the two, one needs to examine 
signatures such as the behaviour of perturbations in the universe. This analysis 
is beyond the scope of our work in this paper.
Finally, in this paper, we have limited our analysis
for the canonical scalar fields. We do not envisage the obvious extension
to the non-standard scalar fields like
non-minimally coupled fields \cite{Amendola:1999qq},
tachyonic fields \cite{Aguirregabiria:2004xd}, or 
phantom fields \cite{UrenaLopez:2005zd} to be complicated.
It may be interesting to consider the dynamics
of such scalar fields as well as multi-field models \cite{Hartong:2006rt}
in curved FRW universes.

\section*{ACKNOWLEDGMENTS}

S. M. is supported by JSPS Postdoctral Fellowships 
for Research Abroad, and M. S. by a University of Nottingham bursary.
E. J. C. is grateful to the Royal Society for financial support.


\end{document}